\documentclass{raa}
\usepackage{amsmath,amsfonts,xfrac,natbib,graphicx}
\newcommand{\cts}{\mathrm{counts/s}}
\newcommand{\vect}[1]{\boldsymbol{#1}}

\newcommand{\ptspercm}{\mathrm{photons/s/cm^2}}
\newcommand{\kev}{\mathrm{keV}}
\newcommand{\crab}{\mathrm{Crab}}
\newcommand{\mcrab}{\mathrm{mCrab}}

\makeatletter
\@namedef{dgt@nnne}{\dg@DX=1 \dg@DY=3 \dg@SIZE=1}
\makeatother

\begin{document}
\title{Point source detection performance of Hard X-ray Modulation Telescope
imaging observation}

\volnopage{Vol.0 (200x) No.0, 000--000}      
\setcounter{page}{1}                         

\author{Zhuoxi Huo\inst{1,2} \and Yiming Li\inst{3} \and Xiaobo Li\inst{5}
\and Jianfeng Zhou\inst{2,4}}
\institute{
1. Department of Physics, Tsinghua University,
Beijing 100084, China;
{\it huozx@tsinghua.edu.cn}\\
2. Centre for Astrophysics, Tsinghua University,
Beijing 100084, China\\
3. Laboratoire de l'Acc\'{e}l\'{e}rateur Lin\'{e}aire,
Orsay 91898, France\\
4. Department of Engineering Physics, Tsinghua University,
Beijing 100084, China\\
5. Institute of High Energy Physics, Chinese Academy of Sciences,
Beijing 100049, China}
\abstract{The Hard X-ray Modulation Telescope (HXMT) will perform an all-sky
survey in hard X-ray band as well as deep imaging of a series of small sky
regions.
We expect various compact objects to be detected in these imaging observations.
Point source detection performance of HXMT imaging observation depends not only
on the instrument but also on its data analysis since images are reconstructed
from HXMT observed data with numeric methods.
Denoising technique plays an import part in HXMT imaging data analysis pipeline
alongside with demodulation and source detection.
In this paper we have implemented several methods for denoising HXMT data and
evaluated the point source detection performances in terms of sensitivities and
location accuracies.
The results show that direct demodulation with $1$-fold cross correlation should
be the default reconstruction and regularization methods, although both
sensitivity and location accuracy could be further imporved by selecting and
tuning numerical methods in data analysis of HXMT imaging observations.
}
\authorrunning{Z.-X. Huo et al.}
\titlerunning{Point source detection performance of HXMT imaging observation}
\maketitle

\section{Introduction}
\subsection{Background}
Hard X-ray Modulation Telescope (HXMT) is a planned Chinese space X-ray
telescope.
The telescope will perform an all-sky survey in hard X-ray band
($1$ -- $250\;\kev$), a series of deep imaging observations at small sky
regions and pointed observations.

We expect a large number of X-ray sources, e.g., AGNs, to be detected in its
all-sky survey.
We also expect through a series of deep imaging observations at the Galactic
plane X-ray transients can be detected\citep{li2007,lu2012}.
Therefore the point source detection performance is one of our concerns on
HXMT data analysis.

Methods and corresponding sensitivities of pointed observation have been
discussed by \citet{jin2010}.
In imaging observations such as all-sky survey and deep imaging at small
sky regions, a variety of data analysis aspects and methods are involved.
\textbf{First, images are computed instead of recorded directly by the optical
instrument.}
Mapping as well as image reconstruction methods are useful.
The direct scientific data products from imaging observations of HXMT are still scientific events, more specifically, X-ray photon arrival events. The attitude control system (ACS) of the spacecraft 
reports the attitudes periodically. We take these reported attitudes as nodes to perform certain interpolations to determine the instantaneous attitude for each scientific event. In this way, a set of 
parameters are assigned to each event, including the coordinates on the celestial sphere where the telescope is pointing at. Hence we call this process \textit{events mapping}, where scientific events 
are mapped from time domain to the celestial sphere. The product of this process is refered to as the observed image, which implies the dimensionality of the data.
\textbf{Second, the exposure to a specific source is limited more strictly}
thus the signal-to-noise ratio (SNR) is tightly restricted.
Hence the importance of regularization methods becomes prominent.
\textbf{Finally, \textit{a picture is worth a thousand words}}.
Various information can be extracted from an image by numerical methods.

In this paper we investigate the point source detection performance of
the imaging and detecting system synthesised from the telescope as well
as diverse combinations of data analysis methods, especially the regularization
methods.

\subsection{Modulation in HXMT imaging observation}
The PSF of HXMT HE, which is a composite telescope consisting of 18 collimated
detectors, describes the response of the telescope to a point source when the                                                                                                                                      
telescope is pointing at the source as well as different locations around it.                                                                                                                                      
In other word, the PSF is a density function of the distribution of responses                                                                                                                                      
occur in different observation states, which is denoted by the instantaneous                                                                                                                                       
attitude of the telescope as a spacecraft.                                                                                                                                                                         
So the PSF takes the attitude of the telescope as its input.                                                                                                                                                       
                                                                                                                                                                                                                   
We use the proper Euler's angles to describe the attitude of the telescope,                                                                                                                                        
i.e., $\phi$ and $\theta$ the longitude and latitude of the pointing, as well                                                                                                                                      
as $\psi$ denoting the rotation angle of the telescope around its own pointing                                                                                                                                     
axis, namely, the position angle.                                                                                                                                                                                  
The modulation equation that corresponds to the imaging observation over the                                                                                                                                       
sphere surface is                                                                                                                                                                                                  
\begin{equation}
d(\phi,\theta,\psi) = \iint\limits_{\Omega} p(\phi,\theta,\psi,\phi',\theta')
  f(\phi',\theta')\cos\theta'\mathrm{d}\phi'\mathrm{d}\theta' \text{,}
\end{equation}
where $f(\phi',\theta')$ is the object function (i.e., the image) defined in
a compact subset of the sphere surface $\Omega$,
$p(\phi,\theta,\psi,\phi',\theta')$ is the modulation kernel function that
relates the value of object function $f(\phi',\theta')$ defined on a
neighbourhood of the point $(\phi',\theta')$ of the subset $\Omega$
to the instantaneous response of the telescope, while its status is 
$(\phi,\theta,\psi)$.

The modulation kernel function is determined by the point spread function (PSF)
of the telescope.
Suppose a unit point source is located at the zenith of the
celestial sphere, i.e., the point $(0,0,1)$ in the corresponding Cartesian
coordinate system.
Fix the position angle $\psi$, while slew the telescope across the polar cap,
and assign responses of the telescope to the unit source to pixels on the
celestial sphere.
Then we obtain a map $P(\phi,\theta)$ represents the PSF on the celestial
sphere with fixed rotation angle $\psi$.

The map is then projected to a tangent plane of the celestial sphere $z=1$ by
gnomonic mapping, i.e.,
\begin{equation}
\left\{
    \begin{aligned}
    u &= \cot\theta\cos\phi \\
    v &= \cot\theta\sin\phi
    \end{aligned}\right.
\label{eq-gnomonic}\text{,}
\end{equation}
where $u$ and $v$ are local Cartesian coordinates on the tangent plane.
Now we have
\begin{equation}
P_{\tan}(u,v) = P_{\tan}(\cot\theta\cos\phi,\cot\theta\sin\phi) = 
  P(\phi,\theta)\text{,}
\end{equation}
i.e., the PSF defined on the tangent plane.

To provide for data analysis we have two discrete forms of $P_{\tan}$,
\begin{equation}
P_{i,j} = \frac{\iint_{\alpha_{i,j}} P_{\tan}(u,v)\mathrm{d}u\mathrm{d}v}
  {\iint_{\alpha_{i,j}} \mathrm{d}u\mathrm{d}v}\text{,}
\end{equation}
where $\alpha_{i,j}$ is the neighbourhood of the pixel $(i,j)$ of the
2-D pixel grid, and the normalized form
\begin{equation}
H_{i,j} = \frac{\iint_{\alpha_{i,j}} P_{\tan}(u,v)\mathrm{d}u\mathrm{d}v}
    {\iint P_{\tan}(u,v)\mathrm{d}u\mathrm{d}v}\text{.}
\label{eq-pixelization}
\end{equation}

Given the discrete image $\vect{F}$, the detection area $A$ and the
duration of exposure on each pixel $\tau$, the observed data is
\begin{equation}
\vect{D} = \tau \cdot \left[\frac{A}{\max\limits_{i,j}H_{i,j}}
    \left(\vect{F} \ast \vect{H}\right) + r_b\right] \text{,}
\label{eq-modulation}
\end{equation}
where $r_b$ is the constant background count rate,
$\ast$ denotes the convolution,
and $\vect{H}$ is the normalized PSF on the tangent plane.
Eq. \ref{eq-modulation} approximates modulations in HXMT imaging observations.

Distortion occurs when projecting a set of points from (or to) a sphere
surface to (or from) a plane. For example, distance between any two
points, area of any continuous subset, angle between any two crossing lines
(or tangents of curves) are altered non-uniformly.
On the other hand, the rotation angle $\psi$ is not always fixed during HXMT
imaging observations.
Both distortions in image reconstruction from HXMT observed data and rotations
during imaging observations have been discussed by \citet{huo2013raa}.
Here for the sake of simplicity, we ignore them in this article.

\section{Numerical methods}
\label{sect-methods}
\subsection{Single point source detection performance estimation}
\label{sect-estimation}
We estimate the single point source performance in terms of sensitivity and
position accuracy through the following procedures.
\begin{enumerate}
\item Determine flux threshold for point source detection.
    \label{item-threshold}
    \begin{enumerate}
    \item Simulate a frame of observed data contains only background counts.
        \label{item-simulate-data}

    \item Run denoise program on the simulated data to try to increase the
    signal-to-noise ratio.
        \label{item-denoise}

    \item Demodulate the denoised data.
        \label{item-demodulate}

    \item Run SExtractor, a source detection program, by
    \citet{bertin1996sextractor} on the demodulated image to detect point
    sources and extract their intensities, coordinates and other parameters.
    At this point a catalog of point sources is compiled from the simulated
    data.
    Point sources detected here, i.e., from images demodulated from background
    data are false detections.
        \label{item-extract-source}

    \item Repeat previous steps (from \ref{item-simulate-data} to
    \ref{item-extract-source}) and a series of catalogs are compiled.
    Draw a histogram of flux of false detections that could possibly been
    detected from background counts given a specific condition of both
    observation and detection.

    \item Choose a cut from the histogram as the flux threshold so that
    a certain percent of the false detections are rejected and the rejection
    percentage is precise enough.
    The rejection percentage, e.g., $95\%$ or $99.7\%$ etc., reflects the
    significance of detections above the corresponding threshold.
    \end{enumerate}

\item Estimate detection efficiency and position accuracy of a point source
of specific flux intensity.
    \label{item-efficiency}
    \begin{enumerate}
    \item Simulate observed data contains a single point source of given
    flux intensity $f_m$ located at $(x_m, y_m)$ in the model image.
        \label{item-simulate-data-pts}

    \item Perform steps from \ref{item-simulate-data} to
    \ref{item-extract-source}.
    A catalog is compiled.
        \label{item-extract-source-pts}

    \item Examine each detection is the catalog.
    Provided the $i-$th source in the catalog is detected at $(x_i, y_i)$
    in the demodulated image, the distance between the extracted source and
    the true point source
    \begin{equation}
      \delta_i = \sqrt{\left(x_m - x_i\right)^2 + \left(y_m - y_i\right)^2}
    \end{equation}
    as well as the flux of the extracted source $f_i$ are investigated to
    determine whether the $i-$th source is a true source or not.
    We define the score of the current catalog in detecting the single point
    source as
    \begin{equation}
        d_k =
        \begin{cases}
            1 & \exists \; i : \left(\delta_i \leq \Delta\right)\land
                \left(f_i \geq F_{thres}\right) \\
            0 & \text{otherwise}
        \end{cases}\text{,}
    \end{equation}
    where $k$ is the index of the current catalog, $\Delta$ and $F_{thres}$
    are position accuracy and flux thresholds,
    therefore the score $d_k$ reveals whether the $k-$th catalog contains
    the true source or not, in other word, through the previous steps
    (simulated observation, denoising, demodulation, source extraction and
    thresholding) whether we have detected the true source effectively.
    If we have, the outcome of these steps is counted as an effective
    detection of the true source, otherwise it is ineffective.
        \label{item-examine-source-pts}

    \item Repeat the previous steps (from \ref{item-simulate-data-pts} to
    \ref{item-examine-source-pts}) $N$ times and calculate the percentage of
    effective detections amongst all detections, namely,
    \begin{equation}
        \eta = \frac{1}{N}\sum_k^N d_k\text{,}
    \end{equation}
    which is defined as the detection efficiency.
    
    Let $(x_k, y_k)$ be the position of the brightest source in the $k-$th
    catalog, the position accuracy is calculated as
    \begin{equation}
    \rho = \frac{1}{\eta N}\sum_k^N d_k \sqrt{\left( x_k - x_m \right)^2 + 
      \left( y_k - y_m \right)^2} \text{.}
    \end{equation}
    \end{enumerate}

\item Find an flux intensity $F_{0.5}$ so that the corresponding detection
effeciency $\eta=50\%$. The intensity $F_{0.5}$ marks the point source
sensitivity of the detecting system synthesised from both the telescope
in specific status and the data analysis program chain.
\end{enumerate}



\subsection{Imaging and mapping}

\subsubsection{Demodulation}
Direct demodulation (DD) method \citep{li1994dd} is used to estimate the
true image from observed data.
Residual map calculated with CLEAN algorithm \citep{hoegbom1974} is used
as lower limit constraint in DD method.
The skewness of the residual map is calculated in each CLEAN iteration and
its minimum absolute value serves as the main stopping criterion of iterations.

\subsubsection{Cross-correlation}
In contrary to detecting and extracting point sources from demodulated images,
it is also feasible to do so from cross-correlated maps as long as the point
sources are isolated with each other, compared to the FWHM of the PSF,
since the position of a peak of the expected value of such a map coincides
with the position of a source regardless of the PSF.
Cross-correlating the observed data and the PSF yields the correlated map
\begin{equation}
\vect{C} = \vect{D} \star \vect{H}\text{,}
\end{equation}
where $star$ denotes cross correlation, $\vect{D}$ and $\vect{H}$ denote the
observed data and the PSF respectively.
The peak of $\vect{C}$ that coincides with a point source of flux intensity
$F$ is
\begin{equation}
C = \frac{\tau \cdot A \cdot F}{\max\limits_{i,j}H_{i,j}}\sum_{i,j}H_{i,j}^2
  \text{,}
\end{equation}
according to Eq. \ref{eq-modulation}.
On the other hand, the background variance of the correlated map is
\begin{equation}
\sigma^2\left(\vect{C}\right) =
  \sum_{i,j}\sigma^2\left(\tau \cdot r_b \cdot H_{i,j}\right) =
  \tau \cdot r_b \cdot \sum_{i,j}H_{i,j}^2\text{,}
\end{equation}
since $\tau \cdot r_b$ follows Poisson distribution.
Hence the significance of the peak in term of numbers of $\sigma$ is
\begin{equation}
\mathrm{SI} = 
  \frac{F \cdot A \cdot \sqrt{\tau \cdot \sum_{i,j}H_{i,j}^2}}
  {\max\limits_{i,j} H_{i,j}\sqrt{r_b}} = 
  \frac{F \cdot A \cdot \sqrt{T}}{\sqrt{r_b}}
    \sqrt{\langle P_{\tan}^2 \rangle}\text{,}
\label{eq-significance}
\end{equation}
where $T$ is total duration of exposure on the 2-D pixel grid,
$\sqrt{\langle P_{\tan}^2 \rangle}$ is the square root of the arithme mean
of $P_{\tan}$ over the 2-D pixel grid, which is determined by the PSF as
well as the range of the pixel grid only, provided the pixel grid is fine
enough (see Eq. \ref{eq-pixelization}).

Cross-correlation significance of isolated point source defined Eq.
\ref{eq-significance} can be evaluated directly, given only the flux intensity
of the source, the background count rate, the duration of exposure,
the detection area and the PSF.
Hence it is determined by the object (i.e., the point source), the telescope
and the status of observation thus effects from data analysis programs are
minimized.

We use the cross-correlation significance as a reference.
For example, in our simulations an isolated point source of $1\;\mcrab$
flux has $2.42\sigma$ significance.

\subsection{Denoising}
\label{sect-denoise-methods}
\subsubsection{Linear methods}
Gaussian smoothing is often used in digital image processing to suppress the
noise at the cost of reduction in resolution.
Trade-off between noise suppression and resolution conservation is adjusted
through the standard deviation $\sigma$ of the Gaussian distribution serving
as the smoothing kernel function.
The best resolution of HXMT HE observed data is about $1.1^\circ$, limited by
the FWHM of its narrow-field collimator.
We set $\sigma$ to $28\arcmin$ so that the FWHM of the Gaussian kernel is
also $1.1^\circ$.

$N-$fold cross correlation transform ($N \ge 1$) can be used in DD method to
regularize the ill-posed problem, more specifically speaking, to ensure the
convergence as well as stability of the solution\citep{li2003dd}.
Here we put this technique in the denoising category.
We have tested $1-$fold and $2-$fold cross correlated DD methods respectively
in this article.

\subsubsection{Non-linear methods}
Non-local means denoising \citep{buades2005non} is an edge-preserving non-linear
denoising method.
To increase its performance we have implemented this method with
fast fourier transforms (FFTs).
The pixel-wisely evaluation of the general Euclidean distance between the $i-$th
pixel and other pixels of an image is replaced by
\begin{equation}
\vect{D}_i = \left\{\sum_k N_{i,k}\cdot W_k + \left(I^2 \ast W\right) - 
  2\left[I \ast \left(N_i \cdot W\right)\right] \right\}^\frac{1}{2}\text{,}
  \label{eq-nlmeans}
\end{equation}
where $N_i$ is neighbourhood of the $i-$th pixel, $N_{i,k}$ is the $k-$ pixel
in the neighbourhood, $I$ is the image, $W$ is weight coefficient of the distance
function.
We use $7 \times 7$ pixels Gaussian kernel with standard deviation $\sigma=2$ (in
pixels) as the weight coefficient $W$ in our simulations.
We reduce the complexity of NLMeans method by computing convolutions in Eq.
\ref{eq-nlmeans} with FFTs instead of using searching windows.
 \citep{buades2008non}.

Median filter is another non-linear edge-preserving denoising method.
This method is effective in removing salt-and-pepper noise in digital images.
In HXMT observed data such noise is incurred typically by missing data or
charged particles of cosmic rays.
We fixed the size of the filter at $2^\circ \times 2^\circ$
(about $50 \times 50$ pixels) in HXMT HE data denoising.

The last non-linear denoising method included in this article is the
adaptive wavelet thresholding with multiresolution support \citep{starck2006}.
The multiresolution support of a noisy image is a subset that contains
significant coefficients only.
So wavelet coefficients that dominated by noise are discarded.
In this article we implemented the non-iterative algorithm.
The $5 \times 5$ $B_3$ spline wavelet is used for multiresolution
decomposition.

\section{Simulation and results}
\label{sect-simulations}
\subsection{In-orbit background simulation}
HXMT HE in-orbit background count rate $r_b$ ranges from $147.6\;\cts$ to
$210.7\;\cts$ \citep{li2009bgrd}.
We use a constant count rate $180\;\cts$ to simulate the average in-orbit
background of HXMT HE in this article.

\subsection{Source energy spectrum and telescope detection efficiency}
We use the formula proposed by \citet{massaro2000crab}  together with
parameters fitted by \citet{jourdain2009crab}
\begin{equation}
F(E) = 3.87 \times E^{-1.79-0.134\log_{10}\left(\sfrac{E}{20}\right)}
\end{equation}
to model energy spectra of Crab-like sources from $20\;\kev$ to $250\;\kev$,
where $E$ is in $\kev$ and the flux $F(E)$ is in $\ptspercm/\kev$.

\begin{figure}[htbp]
\centering
\includegraphics[width=0.6\linewidth]{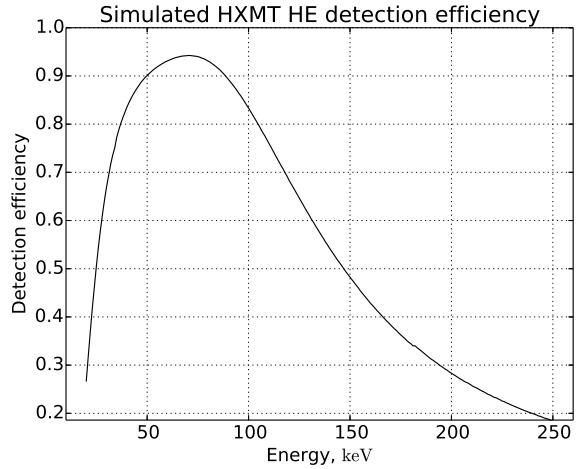}
\caption{HXMT HE detection efficiency}
\label{fig-hxmt-he-eff}
\end{figure}

The  detector efficiency of HXMT HE is derived from its simulated energy
response, as shown in Fig. \ref{fig-hxmt-he-eff}.
Detection efficiency of HXMT HE to a Crab-like source is $67\%$.
Count rate of HXMT HE corresponds to the $1\;\crab$ intensity is $1\,112\;\cts$,
given the detection area of HXMT HE is about $5\,100\;\mathrm{cm^2}$.

\subsection{PSF and modulation}
\begin{figure}[htbp]
\centering
\includegraphics[width=0.6\linewidth]{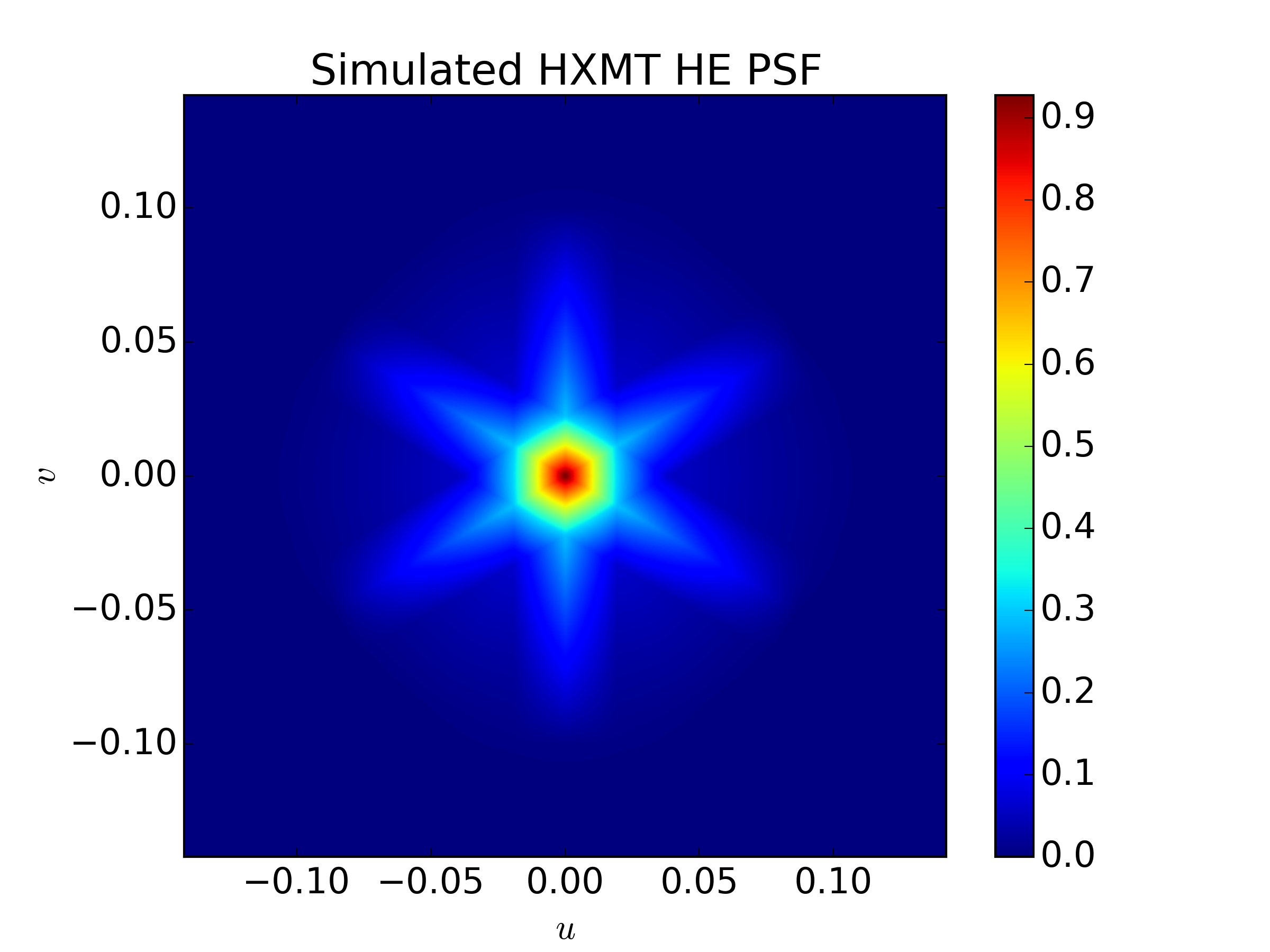}
\caption{Simulated HXMT HE PSF}
\label{fig-hxmt-he-psf}
\end{figure}

We use the diagram shown in Fig. \ref{fig-hxmt-he-psf} to simulate the PSF
$P_{\tan}(u,v)$ on the tangent plane.

\begin{figure}[htbp]
\centering
\includegraphics[width=0.6\linewidth]{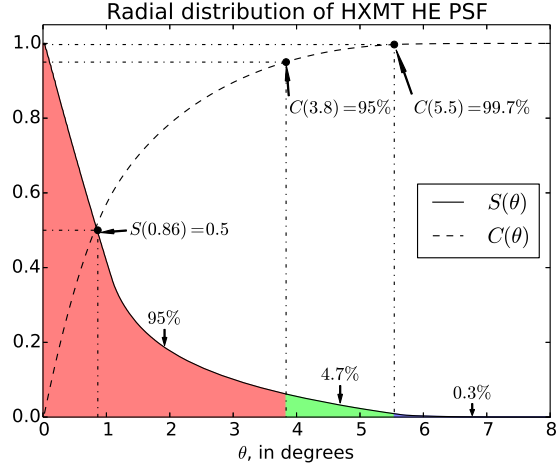}
\caption{Radial distribution of the simulated HXMT HE PSF}
\label{fig-hxmt-he-psf-sl}
\end{figure}

We use the concentric average of $P(\phi,\theta)$, namely,
\begin{equation}
S(\theta) = \frac{\int_{-\pi}^{\pi} P(\phi,\theta)\mathrm{d}\phi}{2\pi}\text{,}
\label{eq-psf-slope}
\end{equation}
and the cumulative sum of $S(\theta)$,
\begin{equation}
C(\theta) = \int_0^{\theta} S\left(\theta'\right)\mathrm{d}\theta'
\end{equation}
to characterize the radial fade-out of the PSF and the concentration of the PSF
respectively, as plotted in Fig.
\ref{fig-hxmt-he-psf-sl}.
From Fig. \ref{fig-hxmt-he-psf-sl} we see that the FWHM of the simulated PSF is
about $1.7^\circ$ while $99.7\%$ responses occur in a diameter of $11^\circ$.

Despite the fact that the direct observed data is scientific events instead of 2-dimensional images, we start our simulation from \emph{simulated} observed data in the form of images defined on 
2-dimensional cartesian pixel grid.
We use a $22^\circ \times 22^\circ$ model image for simulations. Given the
diameter of the PSF, the central $11^\circ \times 11^\circ$ region is fully
modulated, that is, all contributions to observed data in this region are from
the model image only. The surrounding $33^\circ \times 33^\circ$ region is
partially modulated, i.e., part of the contributions to observed data is this
region are from the model image.

The average exposure per unit solid angle is $382\;\mathrm{s/deg^2}$ in HXMT
half-year all-sky survey.
The partially modulated region is discretized by $N \times N$ pixel grid,
so $\tau \approx \sfrac{382 \times 33^2}{N^2}$.
The detection area of each HXMT HE detector is approximately
$300\;\mathrm{cm^2}$ so the total area of all the 17 HXMT HE detectors is
$5\,100\;\mathrm{cm^2}$.

\subsection{Results}
\begin{figure}[htbp]
\centering
\includegraphics[width=0.6\linewidth]{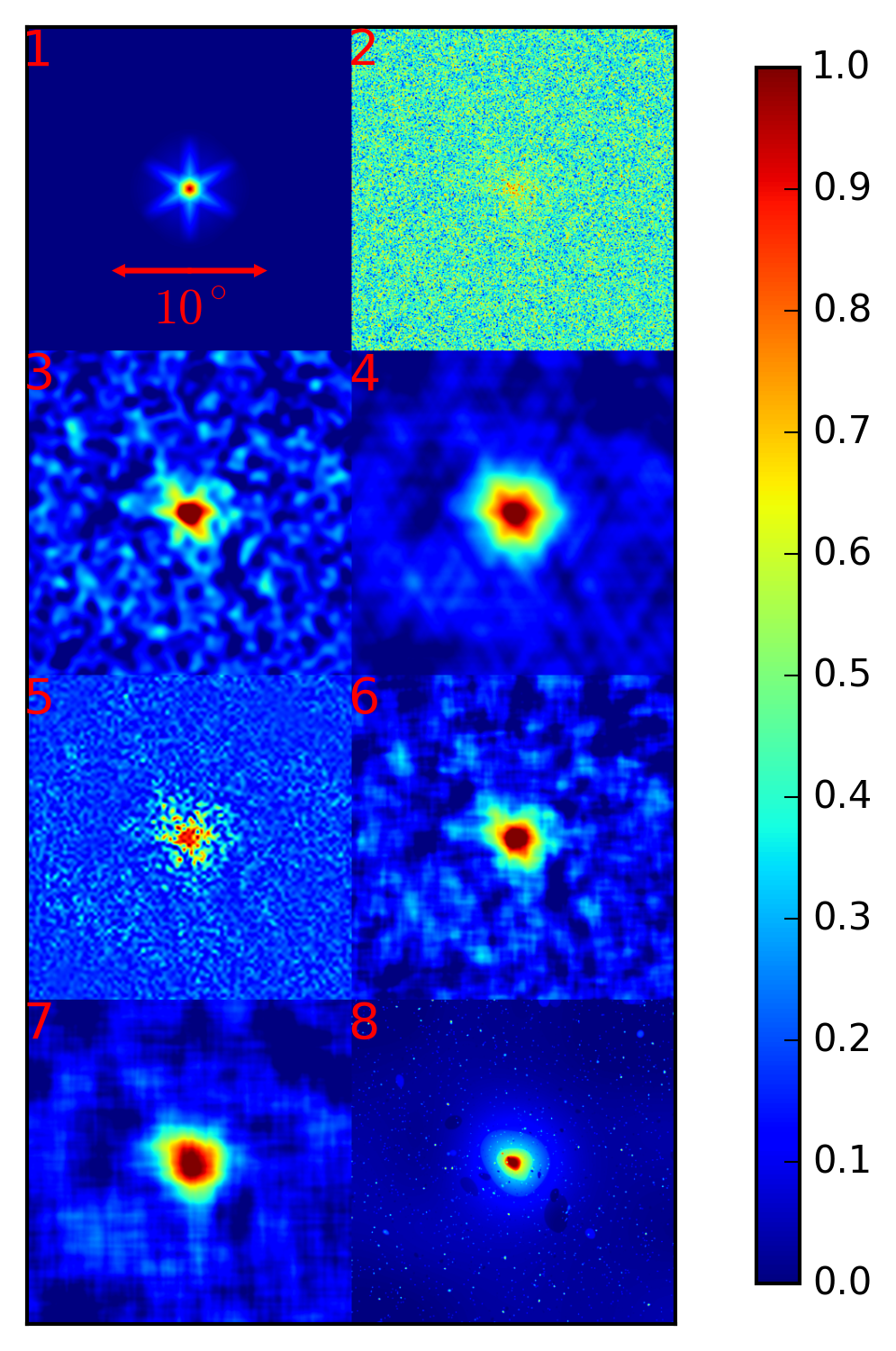}
\caption{Denoising methods. From top to bottom, left to right:\
1. Model image;
2. Observed data of $10\;\mcrab$ point source.
3. Gaussian smoothed data, $\sigma = 28\arcmin$.
4. 1-fold cross correlated data.
5. NLMeans filter denoised data.
6. $2^\circ \times 2^\circ$ median filter denoised data.
7. $4^\circ \times 4^\circ$ median filter denoised data.
8. $3\sigma$ wavelet thresholding denoised data, with $B_3$ spline wavelet transform.}
\label{fig-denoise-methods}
\end{figure}

We have implemented several denoising methods (see Section \ref{sect-denoise-methods}).
Fig. \ref{fig-denoise-methods} shows denoising results by these methods.

We have simulated $5\,000$ frames of observed data that contains the in-orbit
background counts only for each methods to estimate the corresponding flux
thresholds
by the method specified in Procedure \ref{item-threshold} of Section \ref{sect-estimation}.
From false detections we have obtained $2\sigma$ and $3\sigma$ flux thresholds,
see Table \ref{tab-threshold} for  results.

\begin{table}[htbp]
\centering
\begin{tabular}{c|r|r}
Denoising & $2\sigma$ thres., in $\mcrab$ & $3\sigma$ thres., in $\mcrab$ \\ \hline
w/o denoise & $0.520\pm0.001$ & $0.963\pm0.003$\\
1-fold CCT & $0.667\pm0.003$ & $1.192\pm0.010$\\
2-fold CCT & $0.919\pm0.014$ & $1.656\pm0.058$\\
Gaussian, $\sigma = 28\arcmin$ & $0.669\pm0.003$ & $1.133\pm0.008$\\
NLMeans & $0.845\pm0.004$ & $1.340\pm0.011$\\
Median filter, $2^\circ$ & $0.709\pm0.004$ & $1.185\pm0.015$\\
Median filter, $4^\circ$ & $0.665\pm0.005$ & $1.114\pm0.012$\\
Wavelet thres. & $0.0518\pm0.0002$ & $0.216\pm0.002$\\
\end{tabular}
\caption{$2\sigma$ and $3\sigma$ thresholds of point source detection}
\label{tab-threshold}
\end{table}

We have simulated $1\,000$ frames of observd data that contains a Crab-like
point source of $1\;\mcrab$, $1.25\;\mcrab$, $1.5\;\mcrab$, $1.75\;\mcrab$,
$2\;\mcrab$, $2.5\;\mcrab$, $3\;\mcrab$, $4\;\mcrab$, $5\;\mcrab$ and
$10\;\mcrab$ respectively, i.e., $10\,000$ frames of observed data in total.

With these simulated data we have estimated location accuracies as well as
detection efficiencies
by the method described in Procedure \ref{item-efficiency}
of the following methods:
\begin{enumerate}
\item DD without denoising,
\item DD with $1$-fold cross correlation,
\item DD with $2$-fold cross correlation,
\item DD with Gaussian smoothing, where $\sigma=28'$,
\item DD with NLMeans filtering, the size of the filter is
$7 \times 7$ and $\sigma=2$ (both parameters are in pixels),
\item DD with $2^\circ \times 2^\circ$ median filter,
\item DD with $4^\circ \times 4^\circ$ median filter, and
\item DD with adaptive wavelet thresholding.
\end{enumerate}
Implementation details of the above methods are in Sect. \ref{sect-denoise-methods}.

Table \ref{tab-accuracy} shows the location accuracies on simulated data
of $1\;\mcrab$, $2\;\mcrab$, $5\;\mcrab$ and $10\;\mcrab$ point sources.

\begin{table}[htbp]
\centering
\begin{tabular}{c|rrrrrrrrrr}
 & $1\;\mcrab$ & $2\;\mcrab$ & $5\;\mcrab$ & $10\;\mcrab$\\ \hline
w/o denoise & $95 \pm 6$ & $39 \pm 2$ & $9.6 \pm 0.2$ & $5.2 \pm 0.1$ \\
1-fold CCT & $104 \pm 3$ & $53 \pm 1$ & $20.5 \pm 0.4$ & $10.6 \pm 0.2$ \\
2-fold CCT & $149 \pm 5$ & $90 \pm 2$ & $36.9 \pm 0.8$ & $17.2 \pm 0.4$ \\
Gaussian, $\sigma = 28\arcmin$ & $90 \pm 4$ & $38 \pm 1$ & $11.1 \pm 0.2$ & $6.0 \pm 0.1$ \\
NLMeans & $107 \pm 4$ & $52 \pm 1$ & $21.0 \pm 0.3$ & $12.2 \pm 0.2$ \\
Median filter, $2^\circ$ & $91 \pm 4$ & $43 \pm 1$ & $14.3 \pm 0.3$ & $7.7 \pm 0.1$ \\
Median filter, $4^\circ$ & $100 \pm 3$ & $57 \pm 1$ & $23.3 \pm 0.4$ & $14.8 \pm 0.3$ \\
Wavelet thres. & $108 \pm 5$ & $48 \pm 2$ & $11.6 \pm 0.2$ & $5.9 \pm 0.1$ \\
\end{tabular}
\caption{Location accuracies, in arc minutes.}
\label{tab-accuracy}
\end{table}

Table \ref{tab-efficiency} shows the detection efficiencies on simulated data
of $1\;\mcrab$, $1.5\;\mcrab$, $2\;\mcrab$, $2.5\;\mcrab$ and $3\;\mcrab$
point sources.

Although the RL iteration itself we employed in DD method is total-counts-conservative, i.e., the sum of counts of each pixel is conserved after the iteration \citep{richardson1972},
the regularizations including the background constrains as well as various denoising techniques are not necessarily counts-conservative.
As a result, the absolute flux threshold
(Table \ref{tab-threshold})
for rejecting false detections doesn't reflect the sensitivity directly but the detection efficiency
(Table \ref{tab-efficiency})
does.

Errors of flux thresholds, location accuracies and detection efficiencies in Table \ref{tab-threshold},  Table \ref{tab-accuracy} and Table \ref{tab-efficiency} are calculated by bootstrapping. For 
example, following Procedure \ref{item-threshold} a set of $5\,000$ frames of demodulated images are obtained, from which false detections are calculated and a histogram is plotted, where both the 
$2\sigma$ and $3\sigma$ thresholds are determined. Now let's generate a new set with the same volume by resampling from the original set with replacement in order to calculate both the thresholds again. 
We repeat this resampling process until we get enough thresholds calculated to estimate their standard deviations. In Table \ref{tab-threshold}, Table \ref{tab-accuracy} and Table \ref{tab-efficiency} 
each of the errors is a standard deviation that calculated from $1\,000$ resampled sets.

\begin{figure}[htbp]
\centering
\includegraphics[width=0.995\linewidth]{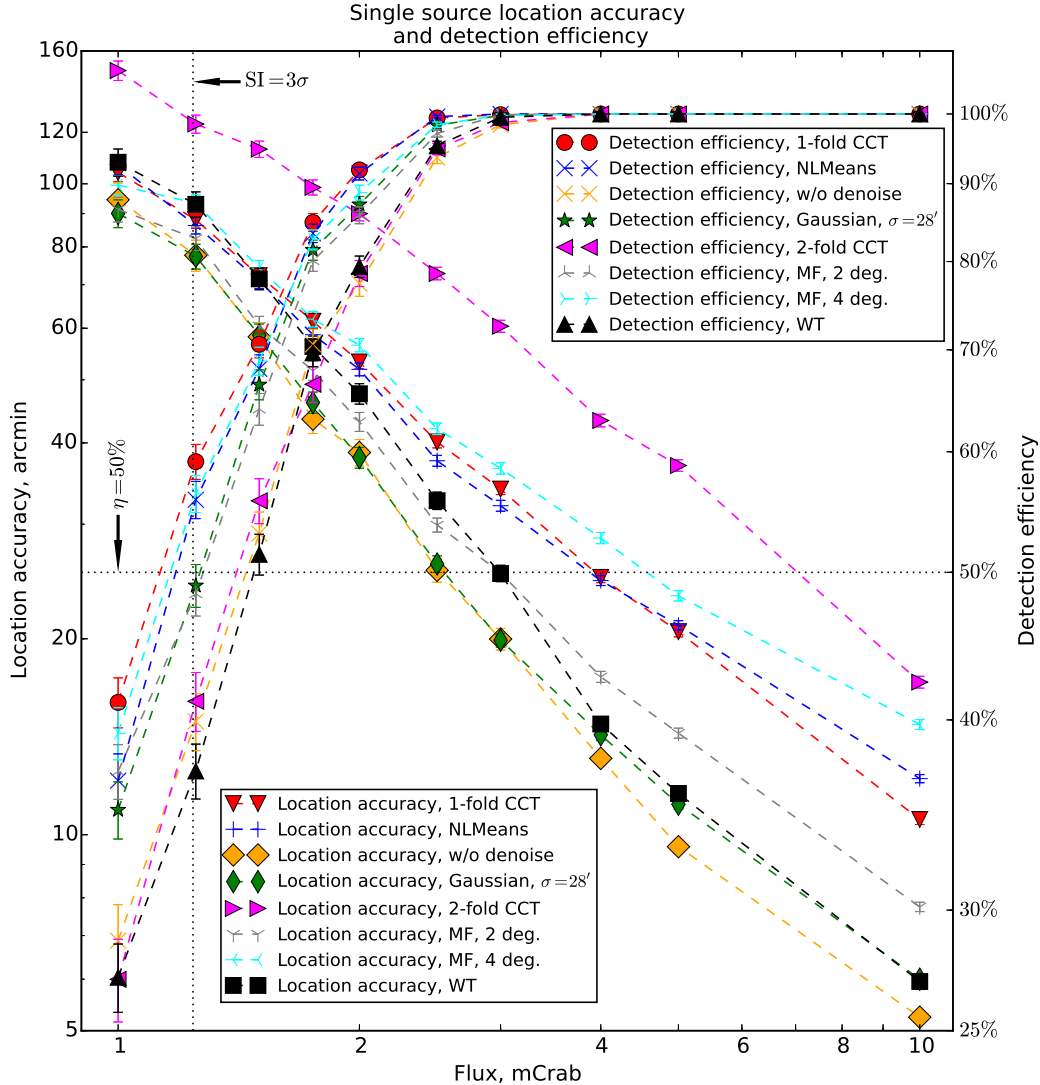}
\caption{Single source location accuracy and detection efficiency.
MF stands for median filter while WT for wavelet thresholding.}
\label{fig-single-source}
\end{figure}

\begin{table}[htbp]
\centering
\begin{tabular}{c|rrrrrrrrrr}
 & $1\;\mcrab$ & $1.5\;\mcrab$ & $2\;\mcrab$ & $2.5\;\mcrab$ & $3\;\mcrab$\\ \hline
w/o denoise & $29 \pm 2$ & $53 \pm 2$ & $77 \pm 1$ & $94 \pm 1$ & $98 \pm 0$ \\
1-fold CCT & $41 \pm 2$ & $71 \pm 1$ & $92 \pm 1$ & $99 \pm 0$ & $100 \pm 0$ \\
2-fold CCT & $27 \pm 2$ & $56 \pm 2$ & $79 \pm 2$ & $95 \pm 1$ & $99 \pm 0$ \\
Gaussian, $\sigma = 28\arcmin$ & $35 \pm 2$ & $66 \pm 2$ & $87 \pm 1$ & $98 \pm 0$ & $100 \pm 0$ \\
NLMeans & $36 \pm 2$ & $68 \pm 1$ & $91 \pm 1$ & $100 \pm 0$ & $100 \pm 0$ \\
Median filter, $2^\circ$ & $37 \pm 2$ & $64 \pm 2$ & $86 \pm 1$ & $97 \pm 1$ & $100 \pm 0$ \\
Median filter, $4^\circ$ & $39 \pm 2$ & $69 \pm 1$ & $89 \pm 1$ & $98 \pm 0$ & $100 \pm 0$ \\
Wavelet thres. & $27 \pm 1$ & $51 \pm 2$ & $79 \pm 1$ & $95 \pm 1$ & $100 \pm 0$ \\
\end{tabular}
\caption{Detection efficiencies, in percentages.}
\label{tab-efficiency}
\end{table}
A comprehensive summary of all the tested methods on all simulated data is
shown in Fig. \ref{fig-single-source}.

\section{Conclusion}
According to the results from our tests no denoising method shows significant
advantage over $1$-fold cross correlated DD method in single point source
detection efficiency.
Therefore it's suggested that $1$-fold cross correlation should be the default
regularization method for single point source detection in HXMT imaging data
analysis.

In the other hand, the location accuracy can be improved with alternative
denoising methods, such as median filter, wavelet thresholding, Gaussian
smoothing with little kernel, or without denoising, according to the results
in this work.

This article is focused on the single point source detection of HXMT imaging
data analysis, where other interesting topics can not be all covered.
Although the alternative denoising methods have been out-performed more or less
by the default $1$-fold cross correlation in their contributions to the detection
efficiency, they have shown certain advantages in location accuracy, and these
features are promising for locating bright transients, multiple sources
resolving, and so on.

\section*{Acknowledgements}
In this work we made use of SciPy\citep{scipy}, PyFITS and AIRE.
PyFITS is a product of the Space Telescope Science Institute, which is operated
by AURA for NASA.
AIRE is a set of computing facilities initiated by Tsinghua Centre for
Astrophysics.
This work was supported by \emph{National Natrual Science Foundation of China}
(NSFC) under grants No. 11373025, No. 11173038 and No. 11403014 as well as
\emph{Tsinghua University Initiative Scientific Research Program} under grant
No. 20111081102.

\bibliographystyle{raa}
\bibliography{main}
\end{document}